\documentclass[3p,times,twocolumn]{elsarticle}

\usepackage{ecrc}

\volume{00}

\firstpage{1}

\journalname{Nuclear and Particle Physics Proceedings}

\runauth{}

\jid{nppp}

\jnltitlelogo{Nuclear and Particle Physics Proceedings}

\usepackage{amssymb}
\usepackage{amsmath}
\usepackage[figuresright]{rotating}

\newcommand{\beq}{\begin{equation}}
\newcommand{\eeq}{\end{equation}}

\newcommand{\rmd}{{\rm d}}
\newcommand{\rme}{{\rm e}}

\newcommand{\mcal}{\mathcal}

\begin{document}

\begin{frontmatter}



\dochead{}

\title{Event-by-event picture for the medium-induced jet evolution}


\author[cea,jy]{Miguel A. Escobedo}
\ead{miguel.a.escobedo-espinosa@jyu.fi}
\author[cea]{Edmond Iancu}

\address[cea]{Institut de Physique Th\'{e}orique, Universit\'{e} Paris Saclay, CNRS, CEA, F-91191, Gif-sur-Yvette, France}
\address [jy]{Department of Physics, P.O. Box 35, 40014 University of Jyv\"{a}skyl\"{a}, Finland}

\begin{abstract}

We discuss the evolution of an energetic jet which propagates through a dense quark-gluon plasma and radiates gluons due to its interactions with the medium. Within perturbative QCD, this evolution can be described as a stochastic branching process, that we have managed to solve exactly.
We present exact, analytic, results for the gluon spectrum (the average gluon distribution) and for the higher n-point functions, which describe correlations and fluctuations. Using these results, we construct the event-by-event picture of the gluon distribution produced via medium-induced gluon branching. In contrast to what happens in a usual QCD cascade in vacuum, the medium-induced branchings are quasi-democratic, with offspring gluons carrying sizable fractions of the energy of their parent parton. We find large fluctuations in the energy loss and in the multiplicity of soft gluons.
The multiplicity distribution is predicted to exhibit KNO (Koba-Nielsen-Olesen) scaling. These predictions can be tested in Pb+Pb collisions at the LHC, via event-by-event measurements of the di-jet asymmetry.

Based on \cite{Escobedo:2016jbm,Escobedo:2016vba}.
\end{abstract}

\begin{keyword}


\end{keyword}

\end{frontmatter}

% \linenumbers

\section{Introduction}
\label{sec:intro}
One of the observables in which the formation of a collective medium in heavy-ion collisions manifests in a very clear way is the dijet asymmetry, the energy difference between two approximately back-to-back jets \cite{Aad:2010bu,Khachatryan:2016erx}. The usual interpretation of this observation is the following:
\begin{itemize}
\item The two energetic jets are initially created in a hard process, due to momentum conservation the two jets will have back-to-back directions and approximately the same energy.
\item For simplicity we consider central collisions in which the interaction plane has rotational symmetry. The point of the collision region in which the hard process takes place is not always the center, a deviation from this point will have as a consequence that the effective size of the medium seen by each jet will be different.
\item The two jets will lose energy inside of the medium, however the amount of energy loss will depend on the size of the medium that they traverse.
\end{itemize}
In summary, the fact that the formation of the two jets does not happen at the center of the fireball translates in an asymmetry in the effective length of the medium seen by each jet that at the same time translates into an asymmetry in the energy loss.

However, this might not be the whole story. In the previous discussion we were assuming that the energy loss is always the same at fixed medium size, in other words, we were neglecting fluctuations. The question is then, how big are these fluctuations? This is one of the main problems we are going to address in this proceedings and the answer we are going to find is that the typical deviation in the energy loss is of the order of the average value and therefore fluctuations can not be neglected. We are going to arrive to this conclusion by performing an analytical computation based on the results obtained in \cite{Blaizot:2013vha,Blaizot:2013hx}. A similar result was obtained recently by a Monte Carlo computation in \cite{Milhano:2015mng}.

%Let us discuss with more detail how fluctuations can affect the dijet asymmetry. Imaging the case in which the hard process happens indeed at the center of the collision and therefore the medium length seen by each jet is exactly the same. Even in this case, because energy loss is a stochastic process, it can happen that by chance one of the jets loses more energy than the other, and if fluctuations are large this will be quite common. In real world there would be fluctuations both in the effective length and in the energy loss and there is no trivial way to disentangle them.
Another issue we want to discuss in this proceedings is what are the event-by-event properties of gluons produced by the energy loss mechanism. 

%In \cite{Blaizot:2013vha,Blaizot:2013hx} it was understood that the energy loss is dominated by the emission of a few quite energetic gluons by the leading particle but these gluons will subsequently branch democratically (in a way that the two daughter gluons have a similar energy) and the result of this secondary branching will also emit more gluons and so on... At the end of the day all the energy that was originally loss by the emission of few gluons will be distributed in a lot of very soft gluons emitted at large angles. This qualitative picture has been confirmed by experimental observations \cite{Chatrchyan:2011sx,Khachatryan:2015lha}. Here we want to understand the event-by-event picture of the production of these soft gluons, this will allow to compare with what is known about the emission of gluons by a jet in the vacuum and the observation of these properties will be a non-trivial cross-check of the energy loss mechanism.

%This work will be divided as follows. In section \ref{sec:jet} we will review the theory that we are going to use to make the computations. In section \ref{sec:average} we are going to review what is already known about the average of the energy loss. In section \ref{sec:2point} we will compute its fluctuations. In section  \ref{sec:npoint} we will study the distribution of soft gluons emitted by a jet losing energy in a medium and we will discover that it fulfils the Koba-Nielsen-Olesen (KNO) scaling \cite{Koba:1972ng}. Finally, in section \ref{sec:concl} we will give our conclusions.
\section{Jet quenching formalism}
\label{sec:jet}
We are going to perform the computation using the BDMPS-Z theory \cite{Baier:1996kr,Zakharov:1996fv}. In this formalism all the information that we need from the medium is encoded in its length $L$ and a parameter called $\hat{q}$ that controls the amount of jet broadening induced by the medium.

% $\hat{q}$ controls jet broadening, the amount of transverse momentum $k_\perp$ that a jet gets due to the random kicks from particles in the medium. Similarly to what happens in a Brownian process $k_\perp\sim\sqrt{\hat{q}L}$.

There are two time-scales that have a very important role in this problem. The first one we are going to discuss is the formation time. 
%In quantum mechanics the precise moment at which a particle branches is not exactly known, the cross-section of any process is the square of the sum of all the amplitudes contributing to that process, therefore there will be contributions from the products of an amplitude and a complex conjugate amplitude in which a given gluon has been emitted at different times. However these two times can not be infinitely far away and their typical difference is called the formation time $\tau_f$. In fact there exists an uncertainty relation that ensures that 
This is given by the uncertainty relation $\tau_f\sim\frac{2\omega}{k_\perp^2}$ where $\omega$ is the energy of the gluon that is being emitted. In a medium $k_\perp\sim\sqrt{\hat{q}\tau_f}$, this gives a self consistent equation that results in $\tau_f\sim\sqrt{\frac{2\omega}{\hat{q}}}$.

Another important time-scale is the branching time. In the BDMPS-Z theory the probability to emit a gluon during a small time $\Delta t$ is
\begin{equation}
P(\omega,\Delta t)\propto\frac{N_c\alpha_s}{\pi}\sqrt{\frac{\hat{q}}{\omega}}\Delta t\,,
\end{equation}
the branching time $\tau_{br}$ is the period after which we are almost sure that a gluon with a given energy will be emitted, looking at the previous equation we can see that $\tau_{br}(\omega)=\frac{\pi}{N_c \alpha_s}\tau_f(\omega)$, this shows that in perturbation theory the formation time is much smaller than the branching time and therefore, at first approximation, the branching process can be thought as an almost classical process in which gluons are formed instantaneously.

The branching time allows to divide the gluons in two different types:
\begin{itemize}
\item Soft gluons have an energy such that $\tau_{br}\ll L$ therefore they will be emitted abundantly. 
%However the soft gluons emitted by the leading particle will not contribute importantly to the energy loss.
\item The harder gluons which are likely to be emitted are those with $\tau_{br}\sim L$, this implies that they will have an energy of order $\omega_{br}\sim\alpha_s^2\hat{q}L^2$. Their emission by the leading particle will dominate the energy loss.
%, in fact we are going to see that the energy loss is of the order of $\omega_{br}$. 
However, this gluons with energy $\omega_{br}$ will subsequently branch and at the end of the day what will be found is a lot of soft gluons emitted at large angles.
\end{itemize}

The equations and the consequences of the multiple branching obtained with the previous assumptions were discussed in \cite{Blaizot:2013vha,Blaizot:2013hx}, there it was observed the importance of the so-called democratic branching, the process in which a parton branches in a way such that the resulting partons have a similar energy. This will be a rare event for the leading particle because their energy is much bigger than $\omega_{br}$, however for the gluons emitted by the leading particle, that will typically have an energy of the order of $\omega_{br}$ or smaller, this will be a very common process and a very efficient way to transfer energy into low energy gluons emitted at large angles. 
%This is completely different to the emission of gluons by a jet in the vacuum in which the emissions tend to be collinear.
\section{The gluon spectrum and the average energy loss}
\label{sec:average}
The main focus of this section is going to be the gluon spectrum that we define as
\begin{equation}
D(x,t)=x\langle\sum_i\delta(x_i-x)\rangle\,,
\end{equation}
where $x$ is the energy fraction carried by the gluon. This quantity evolves with time following the equation \cite{Baier:2000sb}
\begin{equation}
\frac{\partial}{\partial\tau} D(x,\tau)=\int \rmd z \,
{\cal K}(z)\left[\sqrt{\frac{z}{x}}D\left(\frac{x}{z},\tau\right)-\frac{z}{\sqrt{x}}D(x,\tau)\right]\,,
\label{eq:evoD}
\end{equation}
where $\tau=\frac{\alpha_s N_c}{\pi}\sqrt{\frac{\hat{q}}{E}}t=\frac{t}{\tau_{br}(E)}$. $E$ in this case is the energy of the leading particle. The case interesting for jet quenching at LHC is therefore the one in which $\tau\ll 1$, however the case $\tau\sim 1$ is also interesting in order to understand how jets are absorbed by the medium.

%\begin{figure}
%\begin{center}
%\includegraphics[scale=0.4]{./plot_D1.png}
%\end{center}
%\caption{Plot (in log-log scale) of $\sqrt{x} D(x,\tau)$, 
%		with $D(x,\tau)$ given by Eq.~(\ref{Dexact}), as a function of $x$ for various values of $\tau$:
%		solid (black): $\tau=0.1$; dashed--dotted (purple): $\tau=0.2$;
% dashed double--dotted (blue): $\tau=0.4$; dotted (brown): $\tau=0.75$;
%  dashed (green): $\tau=1$; dashed triple--dotted (red) : $\tau=1.35$.
%		}
%		\label{fig:D}
%\end{figure}

The kernel ${\cal K}(z)$ in eq. (\ref{eq:evoD}) has the form
\begin{equation}
{\cal K}(z)=\frac{[1-z(1-z)]^{5/2}}{[z(1-z)]^{3/2}}\,,
\end{equation}
however eq. (\ref{eq:evoD}) has not been analytically solved so far with this kernel. In \cite{Blaizot:2013hx} it was solved with the approximate kernel ${\cal K}_0(z)=\frac{1}{[z(1-z)]^{3/2}}$ and the initial condition $D(x,0)=\delta(1-x)$
\begin{equation}
D(x,\tau)=\frac{\tau}{\sqrt{x}(1-x)^{3/2}}\exp\{-\frac{\pi\tau^2}{1-x}\}\,.
\label{Dexact}
\end{equation}
%In fig. \ref{fig:D} we plot eq. (\ref{Dexact}) at different times, we see that at small times the peak of the leading particle is clearly visible, however as time passes this peak starts to disappear and we see that the energy starts to accumulate in soft gluons. It is also seen that the energy diminishes as time passes, in fact
The previous formula implies that the energy decreases with time, in fact 
\begin{equation}
\langle X(\tau)\rangle=\int_0^1\,dxD(x,\tau)=e^{-\pi\tau^2}\,.
\end{equation} 
This leaves the question of where this missing energy goes. Eq. (\ref{eq:evoD}) is only valid for particles with a momentum much bigger than that of the particles in the medium. This can be quantified by an infrared cut-off in momentum fraction $x_0$, remarkably $D(x,\tau)$ can be accurately computed setting $x_0=0$. The energy that is not captured inside of $D(x,\tau)$ goes to degrees of freedom with energy fraction smaller than $x_0$, in other words, to the medium. In summary, the energy loss into the medium is 
\begin{equation}
\mcal{E}(\tau)=E\left(1-e^{-\pi\tau^2}\right)\,.
\end{equation}
\section{The 2-point function and the fluctuations of the energy loss}
\label{sec:2point}
In order to quantify the importance of the energy loss fluctuations we will compute the variance (more details on the computation are given in \cite{Escobedo:2016jbm})
\begin{equation}
\sigma^2_\mcal{E}=E^2(\langle X^2\rangle -\langle X\rangle^2)\,.
\end{equation}
We already computed the value of $\langle X(\tau)\rangle$. In order to compute $\langle X^2(\tau)\rangle$ apart from $D(x,\tau)$ we also need the 2-point function defined as 
\begin{equation}
D^{(2)}(x,x',t)=xx'\langle\sum_{i\neq j}\delta(x_i-x)\delta(x_j-x')\rangle\,,
\end{equation}
which gives information about the pairs of partons with different energy found inside the jet. Knowing this $\langle X^ 2\rangle$ is determined as
\begin{equation}
\langle X^2(t)\rangle=\int_0^1\,dx xD(x,t)+\int_0^1\,dx\int_0^1\,dx'D^{(2)}(x,x',t)\,.
\end{equation}

$D^{(2)}$ fulfills an evolution equation similar to the one in eq. (\ref{eq:evoD})
%\begin{eqnarray}
%&&\frac{\del }{\del\tau}D^{(2)}(x,x',\tau)=\int \rmd z \,
%  {\cal K}(z)\left[\sqrt{\frac{z}{x}}D^{(2)}\Big(\frac{x}{z}, x',\tau\Big)-\frac{z}{\sqrt{x}}D^{(2)}\big({x},x',\tau\big)\right] + \,\Big( x \,\leftrightarrow\, x'\Big)\nonumber\\
%  &&\qquad +\,\frac{x x'}{(x+x')^2}\,
%   {\cal K}\Big(\frac{x}{x+x'}\Big)\,\frac{1}{\sqrt{x+x'}}\,D(x+x',\tau)\,.
%\label{eq:evoD2}
%  \end{eqnarray}
%The second line in this equations represents the fact that in order to have a pair of partons we need an already existing parton to split (note that our initial condition is $D^{(2)}(x,x',0)=0$). The analogy between the first line of eq. (\ref{eq:evoD2}) and eq. (\ref{eq:evoD}) indicates that after the branching the two resulting partons will evolve independently. Eq. (\ref{eq:evoD2}) has the formal solution
%\begin{equation}
%D^{(2)}(x,x',\tau)=\int_0^\tau\rmd \tau'\int^1_{x}\frac{\rmd x_1}{x_1}\int^{1-x_1}_{x'}\frac{\rmd x_2}{x_2}\,
%      D\bigg(\frac{x}{x_1},\frac{\tau-\tau'}{\sqrt{x_1}}\bigg) D\bigg(\frac{x'}{x_2},\frac{\tau-\tau'}{\sqrt{x_2}}\bigg)
%\,S(x_1,x_2,\tau')\,, \end{equation}
%where $S(x,x',\tau)$ is the second line in eq. (\ref{eq:evoD2}). This solution has a straight-forward interpretation. With $S(x_1,x_2,\tau')$
, this has the solution
% , we are not going to discuss the solution of this equation but we are going to sketch a procedure to obtain it. We compute all the branchings that happen at a time $\tau'$. Then we compute the evolution of each resulting sub-jet from $\tau'$ to $\tau$ independently, for example in the case of the parton with momentum fraction $x_1$ we multiply by $D\bigg(\frac{x}{x_1},\frac{\tau-\tau'}{\sqrt{x_1}}\bigg)$ and integrate for all $x_1$. With this we have the contribution to $D^{(2)}$ from all the branchings that happen at time $\tau'$, to get the full result we just need to integrate for all $\tau'$. %Similarly to what happens with $D$, it is possible to solve this evolution analytically making the approximation ${\cal K}\to{\cal K}_0$
\begin{align}\label{D2exact}
   D^{(2)}(x,x',\tau)\,=\,\frac{1}{2\pi}\frac{1}{\sqrt{x x' (1-x-x')}}\left[\rme^{-\frac{\pi\tau^2}{1-x-x'}}
   -\rme^{-\frac{4\pi\tau^2}{1-x-x'}}\right].
   \end{align}
At small times this equation has a peak corresponding to pairs whose sum of energy correspond approximately to the original energy of the leading particle $x+x'\sim 1$. However this peak will disappear quite quickly as time passes. There is another peak that indicates that there will be a large number of pairs formed by soft particles (both $x$ and $x'$ much smaller than $1$).

%\begin{figure}
%\begin{center}
%\includegraphics[scale=0.4]{./plot_energy.png}
%\end{center}
%\caption{Comparison of the average energy loss with the typical deviation. Both are normalized to $E$}
%\label{fig:plot_energy}
%\end{figure}

With this result we can already compute $\sigma^2_\mcal{E}$.
%\begin{equation}
%\sigma^2_{\mcal{E}}(\tau)=E^2\left(2\pi\tau\big[\text{erf}(\sqrt{\pi}\tau)-\text{erf}(2\sqrt{\pi}\tau)\big]+2\rme^{-\pi\tau^2}
%-\,\rme^{-4\pi\tau^2}-\rme^{-2\pi\tau^2}\right)\,,
%\end{equation}
In the limit $\tau\ll 1$, which is the one interesting for LHC physics%, this turns into
\begin{equation}
\sigma^2_{\mcal{E}}(\tau)=E^2\left(\frac{1}{3}\pi^2\tau^4-\,\frac{11}{15}\pi^3\tau^6\right)+\mathcal{O}(E^2\tau^8)\,,
\end{equation}
this result means that the typical deviation will go like $E\tau^2\sim\omega_{br}$. This means that both the average and the typical deviation are of the same order of magnitude and that both are of the size of $\omega_{br}$. %A comparison between these two quantities is shown in fig. \ref{fig:plot_energy}.

Let us now discuss the phenomenological consequences of this result. We focus in back-to-back pairs from which we assume that initially both of them have an energy $E$ but they will typically see different path length. On top of that we will have the fluctuations of the energy loss mechanism itself that we have just computed, in this situation
\begin{equation}
\langle E_1-E_2\rangle^2=(N_c\alpha_s\hat{q})^2(\langle L_1^2\rangle-\langle L_2^2\rangle)^2\,,
\end{equation}
where the symbol $\langle \cdot\rangle$ applied on $L_x$ means average over the geometry of the fireball in the different events. This equation tells us that the observation of a $\langle E_1-E_2\rangle^2$ different from $0$ indicates an asymmetry in the path length seen by the jets. However, what is observed experimentally is $\langle|E_1-E_2|\rangle$ rather than $\langle E_1-E_2\rangle$. Therefore the following quantity might give a more precise picture of what is actually observed in experiments
\begin{eqnarray}
&&\sigma^2_{E_1-E_2}=\langle (E_1-E_2)^2\rangle-\langle E_1-E_2\rangle^2=\nonumber\\
&&(N_c\alpha_s\hat{q})^2\left[\frac{1}{3}(\langle L_1^4\rangle+\langle L_2^4\rangle)+\sigma^2_{L_1^2}+\sigma^2_{L^2_2}\right]\,,
\label{eq:sigmaexp}
\end{eqnarray}
looking at this equation we see that indeed the asymmetry on the path length contributes but we also see that even in the case $L_1=L_2$ there will be a non-zero contribution. We also see, looking at eq. (\ref{eq:sigmaexp}), that both effects are of the same order of magnitude.
\section{The n-point functions and KNO scaling}
\label{sec:npoint}
In order to compute the average energy loss $\langle \mcal{E}\rangle$ and the average number of particles inside the jet $\langle N\rangle$ we need to know the gluon spectrum $D$. If we want to compute $\langle \mcal{E}^2\rangle$ and $\langle N^2\rangle$ we also need to know $D^{(2)}$. If we want to have more detailed information on the energy loss and the distribution of particles we need to compute higher order n-point functions $D^{(n)}$. They fulfill an evolution equation similar to the one of $D^{(2)}$, they can be analytically solved using the same approximations \cite{Escobedo:2016vba}
\begin{equation}
D^{(n)}(x_1,\cdots,x_n|\tau)=\frac{(n!)^2}{2^{n-1}n}\frac{(1-\sum_{i=1}^nx_i)^{\frac{n-3}{2}}}{\sqrt{x_1\cdots x_n}}h_n\left(\frac{\tau}{\sqrt{1-\sum_{j=1}^nx_j}}\right)\,, 
\label{eq:Dnresult}
\end{equation}
where
\begin{equation}
h_n(l)=\int_0^l\,dl_{n-1}\cdots\int_0^{l_2}\,dl_1(nl-\sum_{i=1}^{n-1}l_i)\rme^{-\pi(nl-\sum_{j=1}^{n-1}l_j)^2}\,.
\end{equation}
%This equation and eq. (\ref{D2exact}) have a very similar physical interpretation. We follow a sequence on $n-1$ branchings at different times (represented by $l_i$) and then we integrate for all the times in which these branchings happened.

As was already mentioned, the interesting limit for LHC is $\tau\ll 1$. If we are also in the limit $x_0\ll \tau^2$ (very small resolution scale) the number of particles will be completely dominated by soft gluons and we can compute the leading order contribution to $\langle N\rangle$ analytically. In the more restrictive case in which $x_0\ll \tau^2$ and also $\pi n^2\tau^2\ll 1$ we can, using eq. (\ref{eq:Dnresult}), do the same for $\langle N^n\rangle$. All the moments of the number of particles will diverge as $x_0\to 0$, however the ratio
\begin{equation}
C_p=\frac{\langle N^p\rangle}{\langle N\rangle^p}=\frac{(p+1)!}{2^p}\,,
\label{eq:Cp}
\end{equation}
will be a constant that does only depend of $p$. This property is called KNO scaling \cite{Koba:1972ng} and appears in several processes in heavy-ion as well as in collider physics. In fact, eq. (\ref{eq:Cp}) corresponds to a negative binomial distribution with parameter $k=2$. This distribution gives the probability of having $n$ successful attempts in a Bernoulli trial before having $k$ failures, in this case $2$. Similar properties were also found in the vacuum \cite{Dokshitzer:1991wu}, there it was seen that KNO scaling is also fulfilled and that the distribution of emitted gluons was approximately described by a negative binomial distribution but this time with $k=3$. In conclusion we can see that the distribution of gluons produced by a jet, either in a medium or in the vacuum, can be approximately described by a negative binomial distribution and therefore they approximately fulfill KNO scaling. The difference is that in a medium fluctuations and correlations are much more important.

%\begin{figure}
%\begin{center}
%\includegraphics[scale=0.4]{./h3plot.pdf}
%\end{center}
%\caption{Comparison of the exact value of $h_3$ with the approximation that give KNO scaling $h_3^0$ and the asymptotic limit for large $l$.}
%\label{fig:h3plot}
%\end{figure}
%We can also obtain an asymptotic result for eq. (\ref{eq:Dnresult}) in the limit $\tau\gg 1$ to complete the analytic knowledge that we can get from this equation
%\begin{equation}
%D^{(n)}(x_1,\cdots,x_n|\tau)=\frac{n!\rme^{-\frac{\pi\tau^2}{1-\sum_{i=1}^n x_i}}(1-\sum_{i=1}^n x_i)^{n-5/2}}{(4\pi)^{n-1}\tau^{n-2}\sqrt{x_1\cdots x_n}}\,,
%\end{equation}
%as a cross-check in fig. \ref{fig:h3plot} we plot the exact result that can be obtained for $h_3$ with the result we obtain in the two asymptotic limit that we have discussed.
\section{Conclusions}
\label{sec:concl}
In this proceedings we have reviewed the computation of the fluctuations of the energy loss. We have seen that they large, of the order of the average value. This means that they can not be neglected when interpreting experimental results. This is particularly important for the dijet asymmetry, our result shows that such an asymmetry can be generated even if the medium path length that each jet traverses is the same. This is in contradiction with the usual picture.

We have also shown that the gluons emitted during the process in which the jet loses energy fulfill KNO scaling and can be approximately described by a negative binomial distribution. Remarkably this is similar to what is found in the vacuum where the physics is very different. Comparing the two cases we see that in the medium correlations and fluctuations are much bigger.
\section*{Acknowledgments}
The  work  of  M.A.E.  has been  supported, during the preparation of the talk and the proceedings, in  part  by  the  European  Research  Council  under  the  Advanced Investigator Grant ERC-AD-267258 and in part by the Academy of Finland, project 303756.



%% References

\nocite{*}
\bibliographystyle{elsarticle-num}
\bibliography{Escobedo_M}






% \end{thebibliography}

\end{document}